\documentclass{PoS}

\usepackage{graphicx,amssymb,amsmath,wasysym,subfigure}
\usepackage[numbers,sort&compress]{natbib}

\title{Status of the Inert Doublet Model of dark matter after Run-1 of the LHC}

\ShortTitle{Status of the Inert Doublet Model of dark matter after Run-1 of the LHC}

\author{\speaker{Andreas Goudelis}\\
        Institute of High Energy Physics, Austrian Academy of Sciences, Nikolsdorfergasse 18, 1050 Vienna, Austria\\
        E-mail: \email{andreas.goudelis@oeaw.ac.at}}

\abstract{The Inert Doublet Model (IDM) is one of the simplest extensions of the Standard Model that can provide a viable dark matter (DM) candidate. Despite its simplicity, it predicts a versatile phenomenology both for cosmology and for the Large Hadron Collider. We briefly summarize the status of searches for IDM dark matter in direct DM detection experiments and the LHC, focusing on the impact of the latter on the model's parameter space. In particular, we discuss the consequences of the Higgs boson discovery as well as those of searches for dileptons accompanied by missing transverse energy during the first LHC Run and comment on the prospects of probing some of the hardest to test regions of the IDM parameter space during the 13 TeV Run.}

\FullConference{The European Physical Society Conference on High Energy Physics\\
		 22-29 July 2015\\
		 Vienna, Austria}

\begin{document}

\section{The Inert Doublet Model}\label{sec:theidm}

The Inert Doublet Model is one of the simplest extensions of the Standard Model (SM). It has been extensively studied in the literature, since despite its simplicity it can affect in a non-trivial manner phenomena as varied as electroweak (EW) symmetry breaking \cite{Deshpande:1977rw,Barbieri:2006dq,Hambye:2007vf,Goudelis:2013uca,Khan:2015ipa,Swiezewska:2015paa}, 
dark matter (DM) \cite{Barbieri:2006dq,Goudelis:2013uca,LopezHonorez:2006gr,Hambye:2009pw,LopezHonorez:2010tb,Gustafsson:2007pc,Agrawal:2008xz,Andreas:2009hj,Nezri:2009jd,Garcia-Cely:2013zga}, 
EW baryogenesis \cite{Chowdhury:2011ga,Borah:2012pu}, 
neutrino masses \cite{Ma:2006km} 
as well as LEP \cite{Lundstrom:2008ai} and LHC phenomenology \cite{Barbieri:2006dq,Goudelis:2013uca,Cao:2007rm,Dolle:2009ft,Miao:2010rg,Gustafsson:2012aj,Arhrib:2012ia,Swiezewska:2012eh,Belanger:2015kga}.

In the IDM, the SM is extended by a second scalar $SU(2)_L$ doublet $\Phi$, which is odd under a new discrete $\mathbb{Z}_2$ symmetry, while all other fields are even. Upon EWSB, the two scalar doublets can be expanded around the vacuum as
\begin{equation}
	H = \left( \begin{array}{c} G^+ \\ \frac{1}{\sqrt{2}}\left(v + h + i G^0\right) \end{array} \right),
	\
	\Phi = \left( \begin{array}{c} H^+\\ \frac{1}{\sqrt{2}}\left(H^0 + i A^0\right) \end{array} \right),
\end{equation}
where $v = \sqrt{2}~\langle 0 | H | 0 \rangle \simeq 246$ GeV is the vacuum expectation value of the neutral component of $H$, $h$ corresponds to the physical SM-like Higgs boson and $G^0/G^{\pm}$ are the Goldstone bosons. The ``inert'' sector consists of a neutral CP-even scalar $H^0$, a pseudo-scalar $A^0$, and a pair of charged scalars $H^{\pm}$. 

Given the symmetries and particle content of the IDM, the only part of the Lagrangian that gets modified at tree-level is the scalar potential which reads
\begin{align}\label{Eq:TreePotential}
	V_0 & = \mu_1^2 |H|^2  + \mu_2^2|\Phi|^2 + \lambda_1 |H|^4+ \lambda_2 |\Phi|^4 \\ \nonumber
		& + \lambda_3 |H|^2| \Phi|^2 + \lambda_4 |H^\dagger\Phi|^2 + \frac{\lambda_5}{2} \Bigl[ (H^\dagger\Phi)^2 + \mathrm{h.c.} \Bigr].
\end{align}
The masses and interactions of the scalar sector can be read off by expanding the fields in \eqref{Eq:TreePotential} in terms of component fields. They are determined by $\lambda_{1\ldots5}$ and $\mu_2$, which can be traded for the physically more intuitive parameter set
\begin{equation}
 	\left\{ m_{h}, \ m_{H^0}, \ m_{A^0}, \ m_{H^{\pm}}, \ \lambda_L, \ \lambda_2 \right\}\,,
	\label{eq:masses}
\end{equation} 
where the Higgs and inert scalar masses are given by
\begin{align}
	m_{h}^2 &= \mu_1^2 + 3 \lambda_1 v^2, \\ 	
	m_{H^0}^2 &= \mu_2^2 + \lambda_L v^2, \label{Eq:mH0tree} \\
	m_{A^0}^2 &= \mu_2^2 + \lambda_S v^2, \\
	m_{H^{\pm}}^2 &= \mu_2^2 + \frac{1}{2} \lambda_3 v^2, 
\end{align}
and the couplings $\lambda_{L,S}$ are defined as
\begin{align}
	\lambda_{L,S} &= \frac{1}{2} \left( \lambda_3 + \lambda_4 \pm \lambda_5 \right)\,,
\end{align}
while $\mu_1^2$ is eliminated through the scalar potential minimization condition $m_{h}^2 = -2\mu_1^2 = 2 \lambda_1 v^2$.

The imposition of the $\mathbb{Z}_2$ symmetry forbids mixing among the components of $H$ and $\Phi$ and has consequences similar to those of $R$-parity in sypersymmetric models. All $\mathbb{Z}_2$-odd particles couple in pairs to the SM ones while the lightest amongst them is stable and, if electrically neutral, can play the role of a DM candidate. The DM phenomenology of the IDM has been studied, \textit{e.g.}, in \cite{LopezHonorez:2006gr,Hambye:2009pw,LopezHonorez:2010tb} and is essentially identical whether $H^0$ or $A^0$ is the lightest $\mathbb{Z}_2$-odd particle (LOP). For simplicity, in what follows we will stick to the hierarchy $m_{H^0} < m_{{A^0}, H^{\pm}}$. 

\section{Constraints and signatures of the IDM}

The IDM is subject to numerous constraints regardless of whether it is treated as a DM model or not (for a recent account \textit{cf} also \cite{Ilnicka:2015jba}). Theoretical consistency requires that the EW vacuum be sufficiently stable and that the scattering matrix be unitary, whereas calculability imposes that all couplings remain perturbative up to some sufficiently large scale. We will impose these requirements up to a scale of at least $10$ TeV unless otherwise stated, following the method of \cite{Goudelis:2013uca} (for the interesting case of relaxing the vacuum stability constraint in order to include the possibility of a sufficiently long-lived metastable vacuum \textit{cf} \cite{Khan:2015ipa,Swiezewska:2015paa}).

Due to the existence of the new scalar doublet, there can be excessive contributions to the oblique parameters $S$, $T$ and $U$, for which we consider the $3\sigma$ ranges from \cite{Baak:2011ze}. The inert scalar masses are then constrained from LEP measurements and in particular from searches for neutralinos and charginos. The recasts of the corresponding searches performed in \cite{Lundstrom:2008ai} and \cite{Pierce:2007ut} respectively yield (assuming $m_{H^0} < m_{A^0}$) the limits $m_{A^0} \gtrsim 100$ GeV and $m_{H^\pm} \gtrsim m_W$. Lastly, for $m_{H^0} \leq m_h/2$, the SM Higgs boson can decay into a pair of $H^0$ particles. The global fit to the Higgs signal strengths performed in \cite{Bernon:2014vta} sets the limit ${\rm BR}(h\rightarrow {\rm inv.}) < 0.12$ at $95\%$ confidence level (CL). The latter constraint restricts the $h-H^0-H^0$ coupling $\lambda_L$ to very small values, $\lambda_L \lesssim 6\times 10^{-3}$.

Once the IDM is considered as a DM model, additional constraints arise from the Planck measurements \cite{Ade:2013zuv} of the DM abundance in the universe as well as from the null searches for DM at the XENON \cite{Angle:2011th,Aprile:2012nq} and LUX experiments \cite{Akerib:2013tjd}. Assuming standard thermal freeze-out, the IDM can reproduce the observed DM abundance in three regimes. In the low-mass regime ($m_{H^0} < m_W$), the LOP pair-annihilates dominantly into $b\bar{b}$ and $\tau^+ \tau^-$ pairs via $s$-channel Higgs exchange, with an increasing contribution from $W^+ W^-$ final states as $m_{H^0} \rightarrow m_W$. In this region the relevant free parameter is the coupling $\lambda_L$ of two DM particles to the Higgs. In the intermediate-mass regime ($m_W < m_{H^0} \lesssim 115$ GeV), annihilation occurs predominantly into $W$ and $Z$ pairs. In a perturbative picture, here annihilation occurs mainly through the interplay of the process involving the $H^0 - H^0 - W^+ - W^-$ coupling, which is a gauge coupling, and the $s$-channel Higgs exchange, which depends on $\lambda_L$. Above a mass of roughly $115$ GeV, the $H^0 - H^0 - W^+ - W^-$ coupling dominates and the predicted relic abundance falls short of the Planck results. However, for $m_{H^0} \gtrsim 500$ GeV, there can be destructive interference among the diagrams involving direct annihilation into $W$, the $s$-channel mediated Higgs exchange and $t$-channel mediated $\mathbb{Z}_2$-odd particle exchange. In this high mass regime, the relic density depends on all parameters of the model.
\begin{figure}[!t]
\begin{center}
\hspace{-1.2cm}
\begin{tabular}{cc} 
\includegraphics[width=0.5\textwidth]{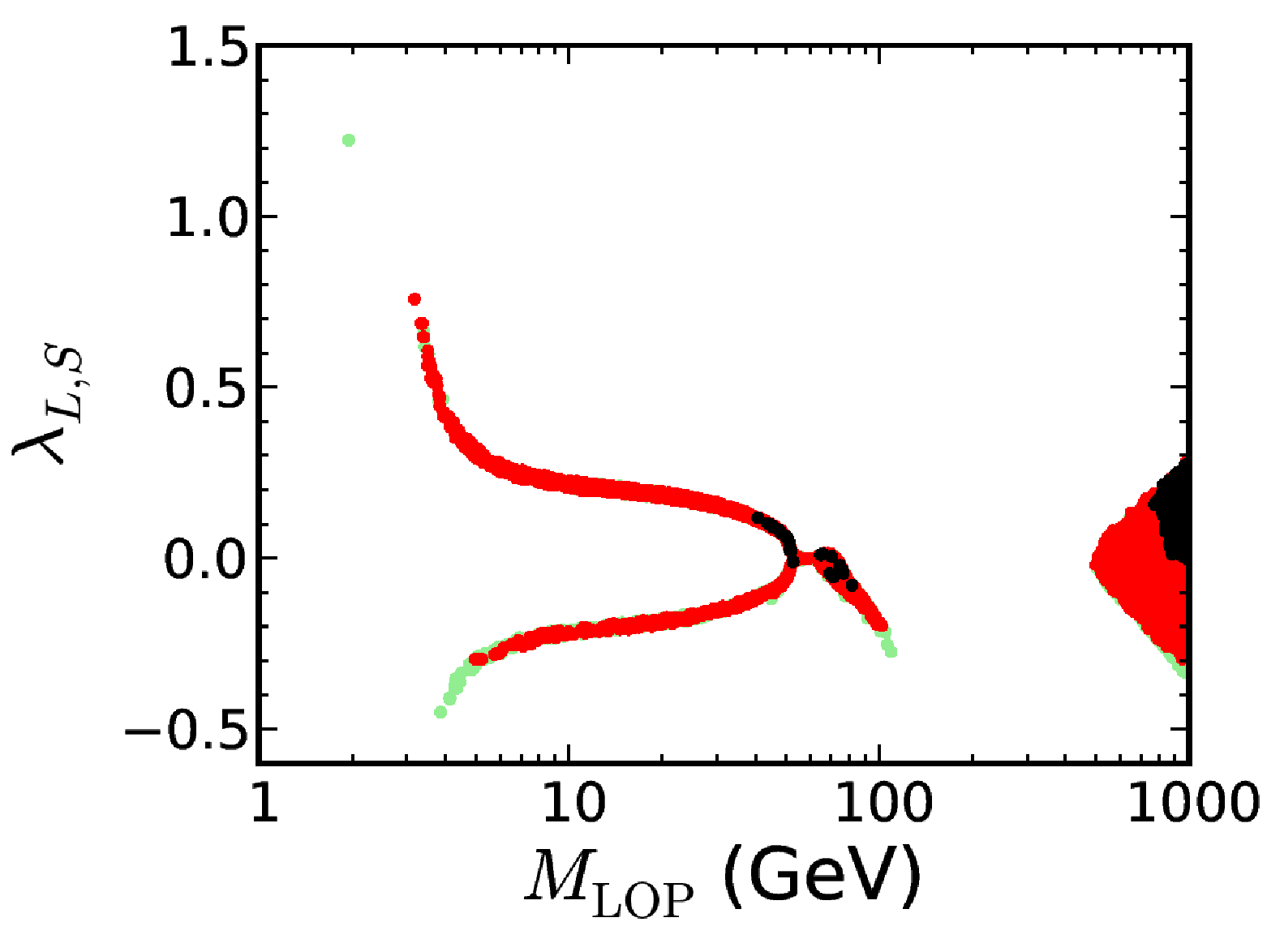}
\includegraphics[width=0.51\textwidth]{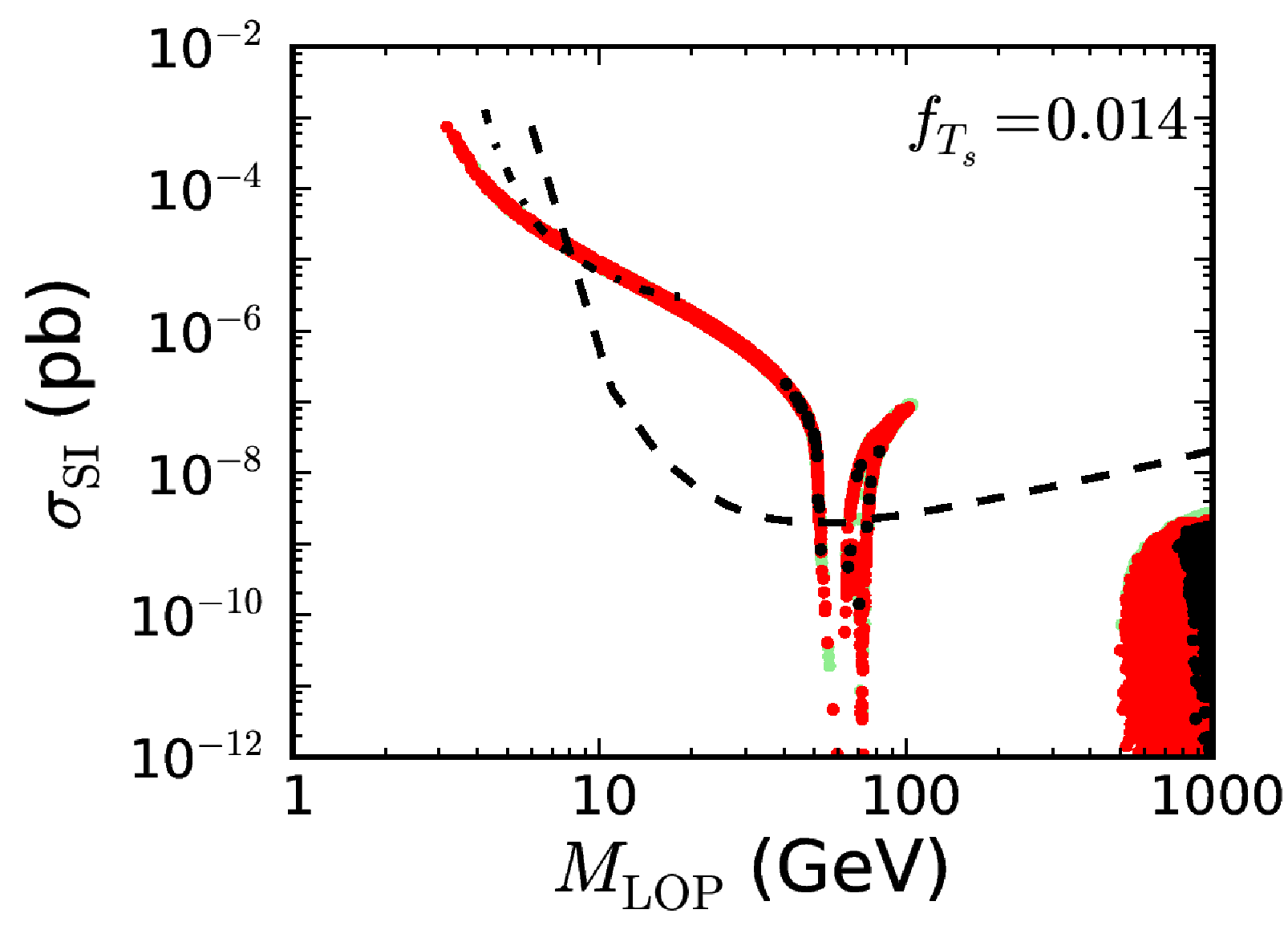}
\end{tabular}
\caption{\textit{(Left panel)} $h$-$H^0$-$H^0$ coupling values $\lambda_L$ satisfying the Planck constraints on the DM abundance in the universe as a function of the dark scalar mass $m_{\rm LOP}$. The differently coloured points correspond to theoretical constraints being satisfied up to different scales, namely $m_Z$ (green), $10$ TeV (red) and $10^{16}$ GeV (black). \textit{(Right panel)} The same, but in the $\sigma_{\rm SI}$ -- $m_{\rm LOP}$ plane. The black lines correspond to the XENON100 (dashed) and XENON10 (dotted-dashed) bounds.}
\label{fig:lamlrelic}
\end{center}
\end{figure}

These remarks are elucidated in the LHS panel of figure \ref{fig:lamlrelic} where we show the values of the $h$-$H^0$-$H^0$ coupling $\lambda_L$ for which both the upper and the lower bound on the dark matter relic density are satisfied as a function of the dark matter particle mass. The $\lambda_L$ values are in $1-1$ correspondence with predictions for the  WIMP-nucleon spin-independent scattering cross-section $\sigma_{\rm SI}$. In the IDM, WIMP-nucleon scattering proceeds solely through $t$-channel Higgs boson exchange and is, hence, fully determined by $\lambda_L$ for a given DM mass. This is shown in the RHS panel of figure \ref{fig:lamlrelic}, where the IDM predictions for the SI scattering cross-section are confronted with the constraints coming from XENON10 and XENON100. Updating these constraints to LUX and CRESST \cite{Angloher:2015ewa} does not qualitatively change this picture, but merely accentuates the tension between the low- and intermediate-mass regime of the model and direct detection constraints. Besides, imposition of the bounds from the invisible Higgs width eliminates all points with masses below roughly $\sim 55$ GeV. 

We can see that in the low- to intermediate mass regime, the only surviving parameter space region is the narrow strip around $m_{H^0} \sim m_h/2$, corresponding to very small values of the $h$-$H^0$-$H^0$ coupling. This is a general feature of numerous DM models: in order to reproduce the observed DM abundance, resonant annihilation in the $s$-channel typically corresponds to a choice of very small couplings, making ``funnel regions'' almost impossible to probe in any experiment. In the IDM case, in the limit $\lambda_{L} \rightarrow 0$ essentially all existing constraints vanish.

\section{Dileptons $+ E_{T}^{\rm miss}$ in the IDM}

The smallness of the $\lambda_L$ coupling after imposition of the invisible Higgs width and direct detection constraints render the IDM funnel region extremely difficult to probe. In \cite{Belanger:2015kga}, we showed that there actually might be hope to constrain this challenging regime, by relying on the production of the \textit{heavier} states of the model leading to a signature of the type $\ell^+\ell^- + E_{T}^{\rm miss}$. There are several ways through which this signature can arise in the IDM, namely
\begin{align} \label{eq:A0H0}
q \bar{q} & \rightarrow Z \rightarrow A^0 H^0 \rightarrow Z^{(*)} H^0 H^0 \rightarrow \ell^+ \ell^- H^0 H^0 , \\ \label{eq:HpHm}
q \bar{q} & \rightarrow Z \rightarrow H^{\pm} H^{\mp} \rightarrow W^{\pm(*)} H^0 W^{\mp(*)} H^0 \\ \nonumber 
          & \rightarrow \nu \ell^+ H^0 \nu \ell^- H^0 , \\ \label{eq:Zh}
q \bar{q} & \rightarrow Z \rightarrow Z h^{(*)} \rightarrow \ell^+ \ell^- H^0 H^0 , \\ \label{eq:fourvertex}
q \bar{q} & \rightarrow Z \rightarrow Z H^0 H^0 \rightarrow \ell^+ \ell^- H^0 H^0 .
\end{align}
Out of these processes, \eqref{eq:Zh} depends on $\lambda_L$ and, given the constraints described in the previous section, is expected to be completely subdominant. The analysis reveals that the signal is in fact dominated by the reaction \eqref{eq:A0H0}.

Two LHC analyses exist that are of relevance for our study. The first is a search for neutralinos, charginos, and sleptons \cite{Aad:2014vma}, leading to the $\ell^+\ell^- + E_{T}^{\rm miss}$ final state through chargino-pair production followed by $\tilde\chi^\pm\to W^{\pm(*)} \tilde\chi^0_1$ or $\tilde\chi^\pm\to \ell^\pm \tilde\nu / \tilde\nu \ell^\pm$ decays, or slepton-pair production followed by $\tilde\ell^\pm\to \ell^\pm \tilde\chi^0_1$ decays. The second is the search for invisible decays of a Higgs boson produced in association with a $Z$ boson \cite{Aad:2014iia}. These two analyses have been recast using the {\sc MadAnalysis 5} \cite{Conte:2012fm,Conte:2014zja} framework. The former, was already available in the Public Analysis Database \cite{Dumont:2014tja} as the recast code \cite{BerangerRecast1}. The latter was implemented and validated \cite{BerangerRecast2} specifically for this work. 

\begin{figure}[!t]
\begin{center}
\includegraphics[width=0.7\textwidth]{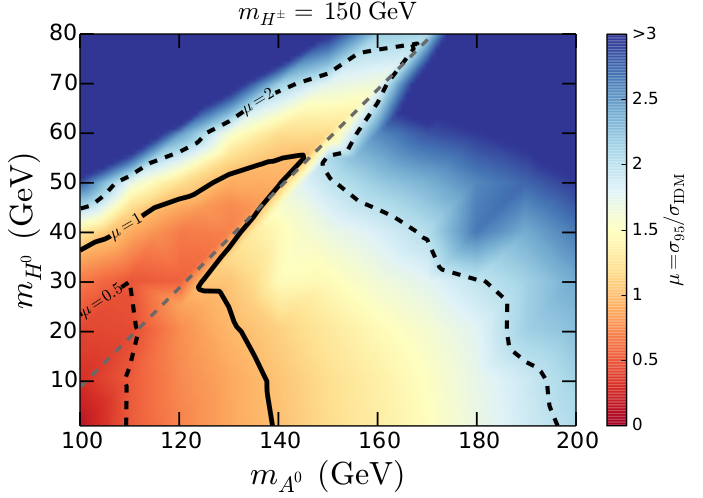}
\caption{The ratio $\mu \equiv \sigma_{95}/\sigma_{\rm IDM}$ in the $(m_{A^0}, m_{H^0})$ plane for the representative value of the charged inert scalar mass $m_{H^\pm} = 150$ GeV. The solid black line depicts the 95\%~CL exclusion contour, $\mu = 1$. The dashed black lines are given for illustration and correspond to the $\mu = 0.5$ and $\mu = 2$ contours. The grey dashed line indicates $m_{A^0}-m_{H^0}=m_Z$.}
\label{fig:dileptonexclusion}
\end{center}
\end{figure}

With the $\lambda_2$ parameter being irrelevant for all observables at tree-level, and given the constraints on $\lambda_L$, we are free to fix the latter to $0$. Besides, $m_{H^\pm}$ is mostly relevant for process \eqref{eq:HpHm} which turned out to be subdominant with respect to \eqref{eq:A0H0}. We then choose two representative values $m_{H^\pm} = 85$ GeV and $m_{H^\pm} = 150$ GeV and scan over $m_{A^0}$ and $m_{H^0}$. Our results are shown in figure \ref{fig:dileptonexclusion} for the case $m_{H^\pm} = 150$ GeV, where we draw contours of the ratio $\mu \equiv \sigma_{95}/\sigma_{\rm IDM}$ in the $(m_{A^0}, m_{H^0})$ plane. Here, $\sigma_{\rm IDM}$ is the cross section predicted by the model while $\sigma_{95}$ is the cross section excluded at 95\%~CL.  With this definition, regions where $\mu \le 1$ are excluded at 95\%~CL.

We can see that the LHC Run-1 excludes $H^0$ masses up to about $35$~GeV for pseudoscalar masses around $100$~GeV, with the limits becoming stronger for larger $m_{A^0}$, reaching $\approx 55$ GeV for $m_{A^0} \approx 145$ GeV. For massless $m_{H^0} \simeq 0$, the LHC excludes $m_{A^0}$ values up to about $\sim$140 GeV. It is interesting that the limits become stronger as the $A^0$ mass increases. This feature is due to the hardening of the lepton spectra (originating from the $A^0\to Z^{(*)}H^0$ decay) with increasing $m_{A^0}$, which in turn results from the opening of phase space amounting to more dileptons passing the signal selection cuts. Besides, when $m_{A^0} \gtrsim m_{H^0} + m_Z$, the $Z$ boson becomes on-shell and the $Z$ veto imposed on the SUSY search channel eliminates the largest portion of the signal. In fact in this region, it is the $Zh\to \ell^+\ell^- + E_T^{\rm miss}$ search that gives the stronger limit. 

For the moment, these constraints seem not to probe any region of the parameter space which could be interesting for DM physics. However, what is important to notice in figure \ref{fig:dileptonexclusion} is that for some $m_{A^0}$ values, the existing limits approach the limit $m_{H^0} \sim m_h/2$, i.e. the funnel region of the IDM. A na\"ive rescaling of signal and background numbers for the LHC $13$ TeV Run (see, \textit{e.g.}, \cite{Bharucha:2013epa}), assuming that the acceptance$\times$efficiency values remain the same, indicates that for an integrated luminosity of 100 fb$^{-1}$, the 95\% CL reach should extend up to $\mu\approx1.2$ (1.6) above (below) the dashed grey line in figure \ref{fig:dileptonexclusion} reaching $\mu\approx2.1$ (2.7) with  300~fb$^{-1}$, thus covering the IDM funnel region. We therefore encourage the experimental collaborations to perform a dedicated search for inert scalars at 13 TeV, which could include not only modified selection cuts but also exploitation of angular separation variables \cite{Dolle:2009ft} in order to further enhance the sensitivity.

\section{Conclusions}

The Inert Doublet Model is a simple (and, hence, predictive) extension of the SM that can accommodate a dark matter candidate. Although the high-mass regime of the model ($m_{H^0} \gtrsim 500$ GeV) is quite difficult to probe in earth-based experiments, its low- and intermediate mass regimes are currently almost excluded by Higgs invisible decay width and direct detection constraints. The only parameter space region currently evading scrutiny is the so-called ``funnel region'', involving annihilation of two DM particles through a resonant $s$-channel Higgs boson (and, to some extent, annihilation into off-shell $W^+ W^-$ pairs). This regime, present in many DM models, is difficult to probe experimentally and could be accessed at the LHC through searches that rely on the production of heavier $\mathbb{Z}_2$-odd states. In particular, searches for dileptons accompanied by missing transverse energy may be able to probe this challenging parameter space region, thus fully excluding the IDM as a DM model for $H^0$ masses below $500$ GeV. This signature could, besides, be exploited in any similar model involving a ``dark'' sector that transforms non-trivially under $SU(2)_L$ and provide interesting constraints for WIMPs of moderate mass.

\bigskip

{\bf Acknowledgements:} \\
AG\ is supported by the New Frontiers program of the Austrian Academy of Sciences and would like to thank the organizers of EPS-HEP 2015 for warm hospitality during the conference.


\begin{thebibliography}{99}

\bibitem{Deshpande:1977rw}
  N.~G.~Deshpande and E.~Ma,
  Phys.\ Rev.\ D {\bf 18} (1978) 2574.
\bibitem{Barbieri:2006dq}
  R.~Barbieri, L.~J.~Hall and V.~S.~Rychkov,
  Phys.\ Rev.\ D {\bf 74} (2006) 015007
  [hep-ph/0603188].
\bibitem{Hambye:2007vf}
  T.~Hambye and M.~H.~G.~Tytgat,
  Phys.\ Lett.\ B {\bf 659} (2008) 651
  [arXiv:0707.0633 [hep-ph]].
\bibitem{Goudelis:2013uca}
  A.~Goudelis, B.~Herrmann and O.~Stål,
  JHEP {\bf 1309} (2013) 106
  [arXiv:1303.3010 [hep-ph]].
\bibitem{Khan:2015ipa}
  N.~Khan and S.~Rakshit,
  Phys.\ Rev.\ D {\bf 92} (2015) 055006
  [arXiv:1503.03085 [hep-ph]].
\bibitem{Swiezewska:2015paa}
  B.~Swiezewska,
  JHEP {\bf 1507} (2015) 118
  [arXiv:1503.07078 [hep-ph]].
\bibitem{LopezHonorez:2006gr}
  L.~Lopez Honorez, E.~Nezri, J.~F.~Oliver and M.~H.~G.~Tytgat,
  JCAP {\bf 0702} (2007) 028
  [hep-ph/0612275].
\bibitem{Hambye:2009pw}
  T.~Hambye, F.-S.~Ling, L.~Lopez Honorez and J.~Rocher,
  JHEP {\bf 0907} (2009) 090
   [JHEP {\bf 1005} (2010) 066]
  [arXiv:0903.4010 [hep-ph]].
\bibitem{LopezHonorez:2010tb}
  L.~Lopez Honorez and C.~E.~Yaguna,
  JCAP {\bf 1101} (2011) 002
  [arXiv:1011.1411 [hep-ph]].
\bibitem{Gustafsson:2007pc}
  M.~Gustafsson, E.~Lundstrom, L.~Bergstrom and J.~Edsjo,
  Phys.\ Rev.\ Lett.\  {\bf 99} (2007) 041301
  [astro-ph/0703512 [ASTRO-PH]].
\bibitem{Agrawal:2008xz}
  P.~Agrawal, E.~M.~Dolle and C.~A.~Krenke,
  Phys.\ Rev.\ D {\bf 79} (2009) 015015
  [arXiv:0811.1798 [hep-ph]].
\bibitem{Andreas:2009hj}
  S.~Andreas, M.~H.~G.~Tytgat and Q.~Swillens,
  JCAP {\bf 0904} (2009) 004
  [arXiv:0901.1750 [hep-ph]].
\bibitem{Nezri:2009jd} 
  E.~Nezri, M.~H.~G.~Tytgat and G.~Vertongen,
  JCAP {\bf 0904}, 014 (2009)
  [arXiv:0901.2556 [hep-ph]].
\bibitem{Garcia-Cely:2013zga}
  C.~Garcia-Cely and A.~Ibarra,
  JCAP {\bf 1309} (2013) 025
  [arXiv:1306.4681 [hep-ph]].
\bibitem{Chowdhury:2011ga}
  T.~A.~Chowdhury, M.~Nemevsek, G.~Senjanovic and Y.~Zhang,
  JCAP {\bf 1202} (2012) 029
  [arXiv:1110.5334 [hep-ph]].
\bibitem{Borah:2012pu}
  D.~Borah and J.~M.~Cline,
  Phys.\ Rev.\ D {\bf 86} (2012) 055001
  [arXiv:1204.4722 [hep-ph]].
\bibitem{Ma:2006km}
  E.~Ma,
  Phys.\ Rev.\ D {\bf 73} (2006) 077301
  [hep-ph/0601225].
\bibitem{Lundstrom:2008ai}
  E.~Lundstrom, M.~Gustafsson and J.~Edsjo,
  Phys.\ Rev.\ D {\bf 79} (2009) 035013
  [arXiv:0810.3924 [hep-ph]].
\bibitem{Cao:2007rm}
  Q.~H.~Cao, E.~Ma and G.~Rajasekaran,
  Phys.\ Rev.\ D {\bf 76} (2007) 095011
  [arXiv:0708.2939 [hep-ph]].
\bibitem{Dolle:2009ft}
  E.~Dolle, X.~Miao, S.~Su and B.~Thomas,
  Phys.\ Rev.\ D {\bf 81} (2010) 035003
  [arXiv:0909.3094 [hep-ph]].
\bibitem{Miao:2010rg}
  X.~Miao, S.~Su and B.~Thomas,
  Phys.\ Rev.\ D {\bf 82} (2010) 035009
  [arXiv:1005.0090 [hep-ph]].
\bibitem{Gustafsson:2012aj}
  M.~Gustafsson, S.~Rydbeck, L.~Lopez-Honorez and E.~Lundstrom,
  Phys.\ Rev.\ D {\bf 86} (2012) 075019
  [arXiv:1206.6316 [hep-ph]].
\bibitem{Arhrib:2012ia}
  A.~Arhrib, R.~Benbrik and N.~Gaur,
  Phys.\ Rev.\ D {\bf 85} (2012) 095021
  [arXiv:1201.2644 [hep-ph]].
\bibitem{Swiezewska:2012eh}
  B.~Swiezewska and M.~Krawczyk,
  Phys.\ Rev.\ D {\bf 88} (2013) 3,  035019
  [arXiv:1212.4100 [hep-ph]].
\bibitem{Belanger:2015kga}
  G.~Belanger, B.~Dumont, A.~Goudelis, B.~Herrmann, S.~Kraml and D.~Sengupta,
  Phys.\ Rev.\ D {\bf 91} (2015) 11,  115011
  [arXiv:1503.07367 [hep-ph]].

\bibitem{Ilnicka:2015jba}
  A.~Ilnicka, M.~Krawczyk and T.~Robens,
  arXiv:1508.01671 [hep-ph].
\bibitem{Baak:2011ze}
  M.~Baak, M.~Goebel, J.~Haller, A.~Hoecker, D.~Ludwig, K.~Moenig, M.~Schott and J.~Stelzer,
  Eur.\ Phys.\ J.\ C {\bf 72} (2012) 2003
  [arXiv:1107.0975 [hep-ph]].
\bibitem{Pierce:2007ut}
  A.~Pierce and J.~Thaler,
  JHEP {\bf 0708} (2007) 026
  [hep-ph/0703056 [HEP-PH]].
\bibitem{Bernon:2014vta}
  J.~Bernon, B.~Dumont and S.~Kraml,
  Phys.\ Rev.\ D {\bf 90} (2014) 071301
  [arXiv:1409.1588 [hep-ph]].
\bibitem{Ade:2013zuv}
  P.~A.~R.~Ade {\it et al.} [Planck Collaboration],
  Astron.\ Astrophys.\  {\bf 571} (2014) A16
  [arXiv:1303.5076 [astro-ph.CO]].
\bibitem{Aprile:2012nq}
  E.~Aprile {\it et al.} [XENON100 Collaboration],
  Phys.\ Rev.\ Lett.\  {\bf 109} (2012) 181301
  [arXiv:1207.5988 [astro-ph.CO]].
\bibitem{Akerib:2013tjd}
  D.~S.~Akerib {\it et al.} [LUX Collaboration],
  Phys.\ Rev.\ Lett.\  {\bf 112} (2014) 091303
  [arXiv:1310.8214 [astro-ph.CO]].
\bibitem{Angle:2011th}
  J.~Angle {\it et al.} [XENON10 Collaboration],
  Phys.\ Rev.\ Lett.\  {\bf 107} (2011) 051301
   [Phys.\ Rev.\ Lett.\  {\bf 110} (2013) 249901]
  [arXiv:1104.3088 [astro-ph.CO]].
\bibitem{Angloher:2015ewa}
  G.~Angloher {\it et al.} [CRESST Collaboration],
  arXiv:1509.01515 [astro-ph.CO].
\bibitem{Aad:2014vma}
  G.~Aad {\it et al.} [ATLAS Collaboration],
  JHEP {\bf 1405} (2014) 071
  [arXiv:1403.5294 [hep-ex]].
\bibitem{Aad:2014iia}
  G.~Aad {\it et al.} [ATLAS Collaboration],
  Phys.\ Rev.\ Lett.\  {\bf 112} (2014) 201802
  [arXiv:1402.3244 [hep-ex]].
\bibitem{Conte:2012fm}
  E.~Conte, B.~Fuks and G.~Serret,
  Comput.\ Phys.\ Commun.\  {\bf 184} (2013) 222
  [arXiv:1206.1599 [hep-ph]].
\bibitem{Conte:2014zja}
  E.~Conte, B.~Dumont, B.~Fuks and C.~Wymant,
  Eur.\ Phys.\ J.\ C {\bf 74} (2014) 10,  3103
  [arXiv:1405.3982 [hep-ph]].
\bibitem{Dumont:2014tja}
  B.~Dumont {\it et al.},
  Eur.\ Phys.\ J.\ C {\bf 75} (2015) 2,  56
  [arXiv:1407.3278 [hep-ph]].
\bibitem{BerangerRecast1}
  B.~Dumont,
  http://doi.org/10.7484/INSPIREHEP.DATA.HLMR.T56W.2
\bibitem{BerangerRecast2}
  B.~Dumont,
  http://doi.org/10.7484/INSPIREHEP.DATA.RT3V.9PJK
\bibitem{Bharucha:2013epa}
  A.~Bharucha, S.~Heinemeyer and F.~von der Pahlen,
  Eur.\ Phys.\ J.\ C {\bf 73} (2013) 11,  2629
  [arXiv:1307.4237].

\end{thebibliography}
\end{document}